\documentclass[twocolumn,showpacs,preprintnumbers, superscriptaddress]{revtex4}

\usepackage{amssymb}
\usepackage{graphicx}
\usepackage{dcolumn}
\usepackage{bm}

\begin{document}

\title{Gap function symmetry and spin dynamics in electron-doped cuprate
superconductor}
\author{C. S. Liu}
\affiliation{Department of Physics, National Taiwan Normal University, Taipei 11650,
Taiwan}
\affiliation{Institute of Theoretical Physics and Interdisciplinary Center of Theoretical
Studies, Chinese Academy of Sciences, P. O. Box 2735, Beijing 100080, China}
\author{W. C. Wu}
\affiliation{Department of Physics, National Taiwan Normal University, Taipei 11650,
Taiwan}

\begin{abstract}
An antiferromagnetic (AF) spin fluctuation induced pairing model
is proposed for the electron-doped cuprate superconductors. It
suggests that, similar to the hole-doped side, the superconducting
gap function is monotonic $d_{x^2-y^2}$-wave and explains why the
observed gap function has a nonmonotonic $d_{x^2-y^2}$-wave
behavior when an AF order is taken into account. Dynamical spin
susceptibility is calculated and shown to be in good agreement
with the experiment. This gives a strong support to the proposed
model.
\end{abstract}

 \maketitle

\section{Introduction}

Pairing symmetry is an important issue towards understanding the
mechanism of superconductivity. For hole-doped cuprate
high-$T_{c}$ superconductor, it is generally accepted that the
pairing symmetry is $d_{x^2-y^2}$-wave \cite{Tsuei969}. On the
other hand, although no consensus has been reached yet, more and
more recent experiments have pointed out that the order parameter
of electron-doped cuprate superconductors is also likely to have a
$d_{x^2-y^2}$-wave pairing symmetry
\cite{Sato2001,Ariando167001,matsui017003,
blumberg107002,qazilbash:214510}. Interestingly however, angle
resolved photoemission measurement (ARPES) \cite{matsui017003} and
Raman scattering \cite{blumberg107002, qazilbash:214510} suggest a
nonmonotonic $d_{x^2-y^2}$-wave order parameter with maxima close
to the nodes (diagonals) rather than to the Brillouin zone (BZ)
boundary. Understanding the origin of the nonmonotonic
$d_{x^2-y^2}$-wave order parameter becomes an important issue.
Yoshimura and Hirashima performed a strong-coupling one-band
calculation and claimed that the nomonotonic feature comes from a
strong AF spin fluctuation \cite{Yoshimura12}. In contrast to
hole-doped cases, they \cite{Yoshimura12} found that the hot
spots, the intersections of the magnetic BZ boundary and the Fermi
surface (FS), are located near the diagonals of the BZ.
Alternatively it was also argued that the nonmonotonic feature of
the order parameter is the outcome of the coexistence of the
superconducting (SC) and the AF orders
\cite{yuan:054501,lu-2006,das:020506}. When AF order coexists with
the SC order, the resulting quasiparticle (QP) excitation can be
gapped by both orders and behave to be nonmonotonic
$d_{x^2-y^2}$-wave, even though the SC gap itself could have a
typical monotonic $d_{x^2-y^2}$ symmetry.

The clue to understand the electron doped cuprate comes from two
doping-dependent FS as revealed by ARPES \cite{Armitage257001,
matsui047005}. These are well explained in terms of the {\bf
k}-dependent band-folding effect associated with an AF order which
splits the band into upper- and lower-branches \cite {kusko140513,
yuan214523}. In the SC state, the QPs could pair each other within
the same band that leads naturally to a two-band/two-gap model.
The two-gap model gives a unified explanation for the upward
feature near $T_c$ and the weak temperature dependence at low $T$
in superfluid density $\rho_s$ \cite{luo027001}. It is also
supported by Hall coefficient and magneto-resistance measurements
\cite{wang1991, jiang1994, fournier1997}.

Raman scattering has the potential to probe different regions of
the FS. It has been shown by Lu and Wang \cite{lu-2006} that SC
and AF orders in electron-doped cuprates are disentangled in Raman
spectra. In our earlier calculation on Raman spectra
\cite{liu:174517}, we have also proved that the Raman shift for
electron-doped cuprates is mainly determined by their pair
breaking associated with different pieces of the FS. The AF order
can cause a vertex correction and enhance the spectral weight, but
nevertheless, it does not change the Raman symmetry. In
particular, near the optimally-doped regime, the frequency of
$B_{2g}$ peak appears to be higher than that of $B_{1g}$ peak
\cite{blumberg107002, qazilbash:214510}. It seems indicating that
the SC gap deviates from the monotonic $d_{x^2-y^2}$-wave.
However, it has been shown that it is indeed two monotonic
$d_{x^2-y^2}$-wave gaps, associated with $\alpha$ and $\beta$-band
FS respectively, which leads to a good description for it
\cite{liu:174517}.

In this paper, spin dynamics is explored to further test the
two-gap model for the electron-doped cuprates. Spin fluctuation is
observable by inelastic neutron scattering (INS), and is confirmed
to be intimately connected with the pairing mechanism in
hole-doped cuprates. In single (CuO$_2$) layer hole-doped cuprates
such as La$_{2-x}$Sr$_x$CuO$_4$, the magnetic peak is always
incommensurate and their incommensurability is robust against the
frequency change \cite{Mason1992, christensen147002}. (Strong
commensurate peak at momentum ${\bf Q}\equiv(\pi,\pi)$ and some
particular resonance frequency $\omega_{\rm r}$ has been observed
in multilayer YBa$_2$Cu$_3$O$_7$ and Bi$_2$Sr$_2$CaCu$_2$O$_8$
though \cite{Bourges1234}.) In current single-layer electron-doped
cuprates such as Nd$_{2-x}$Ce$_x$CuO$_4$ (NCCO), in contrast,
commensurate peak at $\mathbf{Q}$ is observed both in the SC and
normal states \cite{yamada137004, dai:100502, kang:214512,
wilson:157001}. These commensurate peaks survive over a wide
frequency range. It will be shown later that the commensurability
of these magnetic peaks is a natural outcome of the band nesting,
and their robustness is actually incorporated into the existence
of two separate bands. Spin dynamics has been theoretically
examined in various aspects for electron-doped cuprates lately
\cite{tohyama:174517,jianxinli2003,yuan:134522,onufrieva:247003}.

Based on a mechanism making use of the strong AF spin fluctuation,
a pairing model will be proposed for the electron-doped cuprates.
Analogous to the hole-doped side, the SC gap function of the
electron-doped cuprates is thus naturally to have the
$d_{x^2-y^2}$ symmetry in the whole doping range. When the AF
order is significant, it makes a big split between the two bands
and consequently the nonmonotonic $d_{x^2-y^2}$-wave like of the
gap is satisfactorily explained. Of equal importantance, this
model gives a unified picture for the Raman scattering,
$\rho_s(T)$, and INS in electron-doped cuprates.

\section{The Model}

We start with a phenomenological superconducting Hamiltonian
\begin{eqnarray}
H &=&{\sum_{\mathbf{k,}\sigma }}\left[ \varepsilon
_{\mathbf{k}}f_{\mathbf{ k, }\sigma }^{\dagger
}f_{\mathbf{k,}\sigma }+\Delta _{\mathbf{k}}\left( f_{
\mathbf{k},\uparrow }^{\dagger }f_{-\mathbf{k},\downarrow
}^{\dagger }+f_{-
\mathbf{k},\downarrow }f_{\mathbf{k},\uparrow }\right) \right]  \nonumber \\
&&-2Jm{\sum_{\mathbf{k,}\sigma }}^{\prime }\sigma
(f_{\mathbf{k,}\sigma }^{\dagger }f_{\mathbf{k+Q,}\sigma
}+\mathrm{h.c.})-\mu \label{initial Hamiltonian}
\end{eqnarray}
originated from a $t$-$t^{\prime }$-$t^{\prime \prime }$-$J$
model. The slave-boson transformation and spin-density-wave
mean-field approximation are undertaken. Here $f_{\mathbf{k,}
\sigma }^{\dagger }$ ($f_{\mathbf{k,}\sigma }$) is the fermionic
spinon creation (destruction) operator, $m=(-1)^{i}\langle
S_{i}^{z}\rangle $ is the AF order, $\mu$ is the chemical
potential, and
\begin{eqnarray}
\varepsilon _{\mathbf{k}} &=&(2|t|\delta -J\chi )(\cos k_{x}+\cos
k_{y})-4t^{\prime }\delta \cos k_{x}\cos k_{y}  \nonumber \\
&&-2t^{\prime \prime }\delta (\cos 2k_{x}+\cos 2k_{y})
\label{dispersion}
\end{eqnarray}
is the independent particle dispersion with $\delta $ the doping
concentration and $\chi =\langle f_{i\sigma }^{\dagger }f_{j\sigma
}\rangle $ the uniform bond order. The prime denotes that momentum
summation is over the magnetic BZ only ($-\pi \leq k_{x}\pm
k_{y}\leq \pi $). The SC gap function is given self-consistently
\begin{eqnarray}
\Delta _{\mathbf{k}} ={\sum_{\mathbf{k}^\prime}}V( \mathbf{ k,k}
^{\prime }) \langle f_{\mathbf{k}^{\prime },\uparrow }^{\dag }f_{-
\mathbf{k}^{\prime },\downarrow }^{\dag }\rangle, \label{gap
equation0}
\end{eqnarray}
where $V( \mathbf{k,k}^{\prime })$ is the pairing potential. Using
the unitary transformation
\begin{equation}
\left(
\begin{array}{l}
f_{\mathbf{k,}\sigma } \\
f_{\mathbf{k+Q,}\sigma }
\end{array}
\right) =\left(
\begin{array}{cc}
\cos \theta _{\mathbf{k}} & \sigma \sin \theta _{\mathbf{k}} \\
\bar{\sigma}\sin \theta _{\mathbf{k}} & \cos \theta _{\mathbf{k}}
\end{array}
\right) \left(
\begin{array}{l}
\alpha _{\mathbf{k,}\sigma } \\
\beta _{\mathbf{k,}\sigma }
\end{array}
\right)  \label{unitary transformations}
\end{equation}
with $\theta_{\mathbf{k}}$ being defined by $\tan 2\theta
_{\mathbf{k}}=4Jm/[\sigma (\varepsilon _{\mathbf{k+Q}
}-\varepsilon _{\mathbf{k}})]$, Hamiltonian (\ref{initial
Hamiltonian}) can be transformed as \cite{nazario144513,
Das0604213}
\begin{equation}
H={\sum_{\mathbf{k}\sigma l}}^{\prime }\left[ \xi
_{\mathbf{k}l}l_{\mathbf{k} \sigma }^{\dagger }l_{\mathbf{k}\sigma
}+\Delta _{\mathbf{k}}(l_{\mathbf{k} \uparrow }^{\dag
}l_{-\mathbf{k}\downarrow }^{\dag }+l_{-\mathbf{k}\uparrow
}l_{\mathbf{k}\downarrow })\right] -\mu _{l},  \label{two band
model}
\end{equation}
where $l\equiv \alpha ,\beta$, $\mu_\alpha+\mu_\beta=\mu$, and
\begin{eqnarray}
\xi _{\mathbf{k}l}=\frac{\varepsilon _{\mathbf{k}}+\varepsilon
_{\mathbf{ k+Q }}}{2}\mp \sqrt{\frac{(\varepsilon
_{\mathbf{k+Q}}-\varepsilon _{\mathbf{ k} })^{2}}{4}+4J^{2}m^{2}}
\label{2band}
\end{eqnarray}
corresponding to the QP (with only the AF order in it) dispersions
of the two ($\alpha$ and $\beta$) bands.

Throughout this paper, $|t|=0.326$ eV is taken as the energy unit
together with $t^{\prime }=0.3,t^{\prime \prime }=-0.2$, and
$J=0.3$. Other doping-dependent parameters which agree well with
the band FS are listed in Table~\ref{parameters}.

\begin{figure}[tbp]
\vspace{-3.2cm}\includegraphics[width=0.4\textwidth]{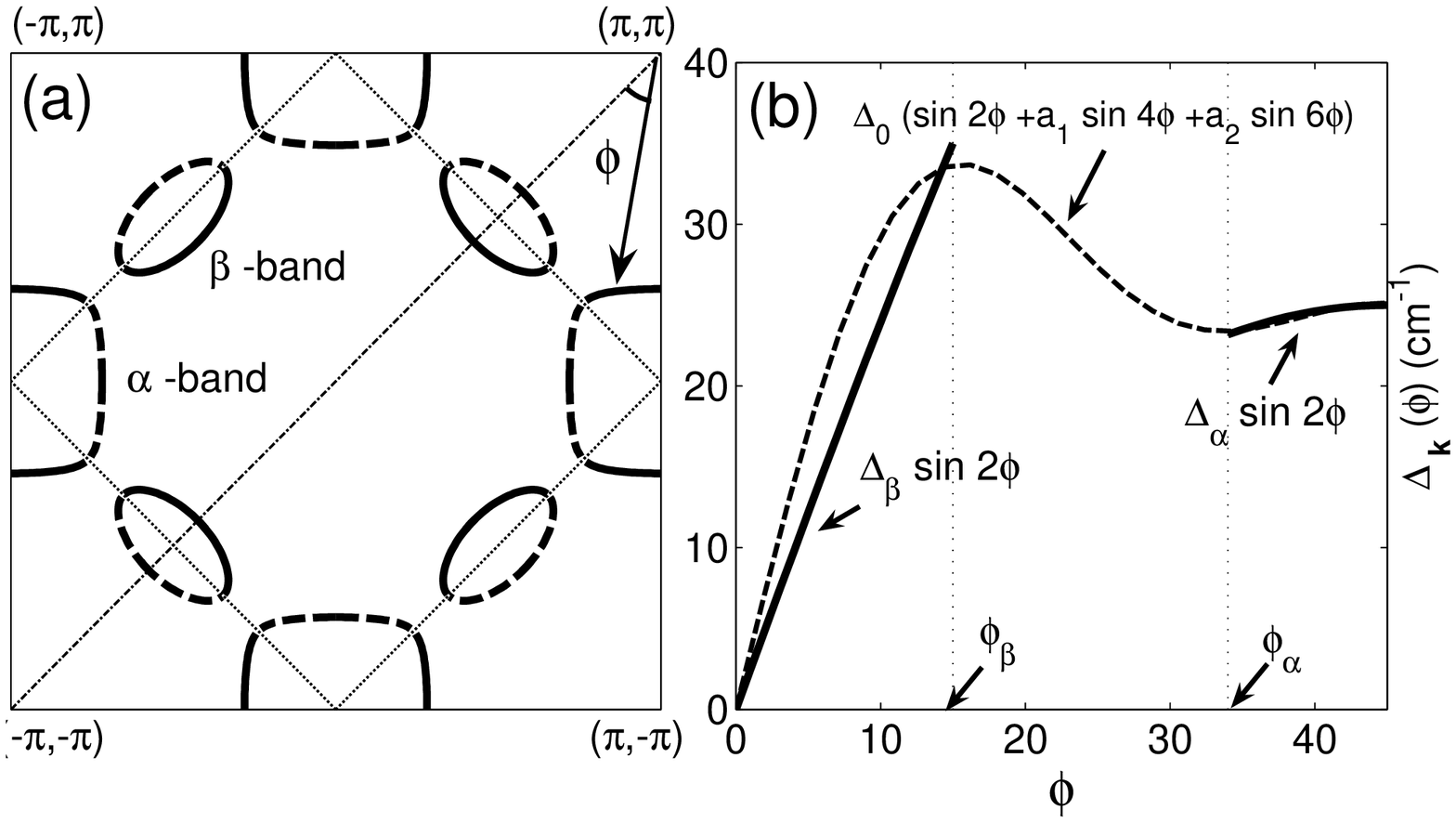}

\vspace{-3.8cm}
\includegraphics[width=0.45\textwidth]{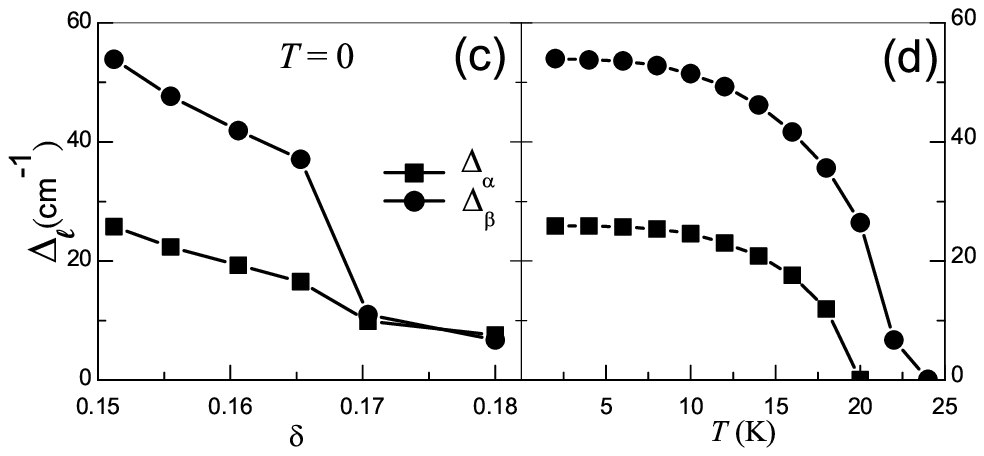}
\vspace{-0.0cm} \caption{(a) Separate $\alpha$- and $\beta$-band
FSs of the electron-doped cuprates with an AF order. (b) Gap
function $\Delta_{\bf k}(\phi)$ with $\phi$ measured along the FS.
Dash line: a nonmonotonic $d_{x^2-y^2}$-wave $\Delta_\mathbf{k}
=\Delta_0 \protect\gamma_ \mathbf{k}$ with
$\protect\gamma_\mathbf{k}=\sin 2 \protect\phi + a_1 \sin 4
\protect\phi + a_2 \sin 6\protect\phi$ proposed in
Ref.~\protect\cite{venturini149701}. Here $\Delta_0 = 33.5$
$\mathrm{cm}^{-1}$, $a_1=0.42$, and $a_2=0.17$. Solid lines: the
piecewise monotonic $d_{x^2-y^2}$-wave $\Delta_\mathbf{k}
=\Delta_l \protect\gamma_\mathbf{k}$ with
$\protect\gamma_\mathbf{k}=\sin 2 \protect \phi$ proposed by us.
For $0 < \protect \phi \leq \protect\phi_\protect\protect\beta$,
$\Delta_l\equiv\Delta_\beta = 70$ $\mathrm{cm}^{-1}$, while for
$\protect\phi_\protect\protect\alpha \leq \protect\phi <
\protect\pi/4$, $\Delta_l\equiv \Delta_\alpha = 25$ $
\mathrm{cm}^{-1}$. The two gap amplitudes,
$\Delta_\protect\protect\alpha$ and
$\Delta_\protect\protect\beta$, are calculated using
(\ref{self-consistent gap equation}) as a function of doping (c)
and temperature (d). See Table~I for parameters.} \label{fig1}
\end{figure}

Fig.~\ref{fig1}(a) shows the FSs for a typical optimally-doped
sample. The AF correlation splits the continuum FS into two pieces
of sheet, in which $\alpha $ band is crossed by the Fermi level in
the antinodal region, while $\beta $ band is crossed by the Fermi
level in the nodal region. Correspondingly small FS pockets appear
around $(\pm \pi ,0)$ and $(0,\pm \pi )$ for the $\alpha$ band,
while separate FS pockets appear centered at $(\pm \pi {/2},\pm
\pi {/2})$ for the $\beta$ band.

\section{Gap symmetry: monotonic vs. nonmonotonic
$d_{x^2-y^2}$-wave}

As mentioned previously, ARPES \cite{matsui017003} and Raman
\cite{blumberg107002} reveal the gap function $\Delta
_{\mathbf{k}}$ in electron-doped cuprate SC to have a
non-monotonic $d_{x^2-y^2}$-wave like. For example in
Ref.~\cite{venturini149701}, $\Delta _{\mathbf{k}}\equiv
\Delta_0\gamma_{\bf k}$ with $\gamma _{\mathbf{k}}=\sin 2\phi
+a_{1}\sin 4\phi +a_{2}\sin 6\phi$ ($\phi$ being the angle
measured relative to the diagonal on the FS) was used to simulate
a non-monotonic $d_{x^2-y^2}$-wave gap [see also
Fig.~\ref{fig1}(b)]. In this kind of approaches, parameters
$a_{1}$ and $a_{2}$ are doping dependent.

As far as SC gap is concerned, it is physically more appealing
that $\gamma _{\mathbf{k}}$ remains the same for the entire doping
range, so long as the pairing mechanism remains the same (no
quantum criticality occurs). Based on a mechanism induced by the
strong AF spin fluctuation, we propose the following piecewise
model for the pairing potential
\begin{equation}
V( \mathbf{k,k}^{\prime }) =\left\{
\begin{array}{lc}
g_{\alpha }\gamma _{\mathbf{k}}\gamma _{\mathbf{k}^{\prime }}, &
\mathrm{~~for~}
\mathbf{k,k^{\prime }}\text{ on }\alpha \text{-band FS}, \\
g_{\beta }\gamma _{\mathbf{k}}\gamma _{\mathbf{k}^{\prime }}, &
\mathrm{~~for~}
\mathbf{k,k^{\prime }}\text{ on }\beta \text{-band FS}, \\
0, & \mathrm{~~otherwise},
\end{array}
\right.  \label{pairV}
\end{equation}
where $\gamma _{\mathbf{k}}=\sin 2\phi$, and $g_\alpha$ and
$g_\beta$ are the two coupling constants. Inspired by the
hole-doped side, it is promisingly to have $\gamma _{\mathbf{k}}$
having the monotonic $d_{x^2-y^2}$ symmetry. This is strongly
supported by the INS experiment and will be elaborated later.
Nevertheless, there are two $d_{x^2-y^2}$-wave gaps, possibly with
different amplitude, for the current electron-doped side. When AF
order breaks down with increasing the doping, the two bands
[Eq.~(\ref{2band})] will eventually merge into a single one. In
this regime, the behavior of the electron-doped cuprates is
expected to be very similar to that of the hole-doped ones, with
one {\em single} monotonic $d_{x^2-y^2}$-wave gap
\cite{liu:174517}. The latter is confirmed by the Raman experiment
\cite{qazilbash:214510}.

The piecewise feature in $V( \mathbf{k,k}^{\prime })$ comes
naturally for the AF spin fluctuation induced mechanism (see
Sec.~IV). Only QPs within the same band favor the pairing
associated with the ${\bf Q}$ wavevector . This is also supported
by the superfluid data which unambiguously reveals that QPs in
$\alpha$ (or $\beta$) band FS pair each other to form SC QPs. But
no SC QP forms in the region where FS is absent \cite{luo027001}.

\begin{table}[tbp]
\caption{Parameters used to calculate $\Delta_\alpha$ and
$\Delta_\beta$.} \label{parameters}
\begin{ruledtabular}
\begin{tabular}{lcccr}
$\delta $ & $m$ & $-\chi $ & $-\mu _{\alpha }$ & $\mu _{\beta }$
\\ \hline 0.15 & 0.178 & 0.15 & 0.005 & 0.078 \\
0.155 & 0.169 & 0.16 & 0.007 & 0.079 \\  0.16 & 0.160 & 0.17 &
0.008 & 0.079 \\  0.165 & 0.150 & 0.18 & 0.009
& 0.079 \\  0.17 & 0.040 & 0.20 & -0.040 & -0.036 \\
 0.18 & 0.010 & 0.20 & -0.034 & -0.035 \\
\end{tabular}
\end{ruledtabular}
\end{table}

Substitution of (\ref{pairV}) and (\ref{unitary transformations})
into (\ref{gap equation0}), one obtains the self-consistent gap
equation respectively for each band
\begin{equation}
1=g_{l}{\sum_{\mathbf{k\in }l~{\rm FS}}}^{\prime }\frac{\gamma
_{\mathbf{k}}^{2}}{2E_{ \mathbf{k}l}}\tanh
\left(\frac{E_{\mathbf{k}l}}{2k_{B}T}\right),
\label{self-consistent gap equation}
\end{equation}
where $E_{\mathbf{k}l}=\sqrt{\xi _{\mathbf{k}l}^{2}+\Delta
_{\mathbf{k}l}^{2}}$ and $\Delta_{{\bf k}l}=\Delta_l \gamma
_{\mathbf{k}}$. In practice, the $\mathbf{k}$ sum in
(\ref{self-consistent gap equation}) can be effectively extended
to the whole MBZ because contribution due to the ${\bf k}$ points
distant from the corresponding $l$-band FS is negligible.
Fig.~\ref{fig1}(b) shows an example of piecewise monotonic
$d_{x^2-y^2}$-wave $\Delta_{{\bf k}l}$, compared with a
nomonotonic one. The region between $(\phi_\beta,\phi_\alpha)$ is
where FS is absent and no SC gap associated with.

Shown in Fig.~\ref{fig1}(c) are the two gap amplitudes,
$\Delta_\alpha$ and $\Delta_\beta$, calculated at $T=0$ with
various doping levels. As the doping $\delta$ decreases (and hence
the AF order $m$ increases), the ratio of
$\Delta_\beta/\Delta_\alpha$ increases along with the gapped
$(\phi_\beta,\phi_\alpha)$ region opens up. This manifests the
nonmonotonic $d_{x^2-y^2}$-wave like gap nicely. Both
$\Delta_\alpha$ and $\Delta_\beta$ decrease as doping increases.
At over doping ($\delta \geq 0.17$), $m$ approaches zero and FSs
join to one piece, $\Delta_\alpha$ and $\Delta_\beta$ match. These
consistent results give strong support to the model pairing (\ref
{pairV}). The parameters used are listed in
Table~\ref{parameters}. Coupling constants, $g_{\alpha}$ and
$g_{\beta}$, are fixed at 0.34 and 0.62 respectively, that give
the best fit for optimal doping ($\delta=0.15$).
Fig.~\ref{fig1}(d) displays the temperature dependence of
$\Delta_\alpha$ and $\Delta_\beta$ ($\delta=0.15$). The SC $T_c$,
determined by the higher of the onset temperatures that make
$\Delta_\alpha$ or $\Delta_\beta$ vanish, is found to be about
$25$ $\mathrm{K}$. The two onset temperatures, differed by 4K or
so, are in good agreement with the upward curvature observed in
$\rho_{s}$ near $T_{c}$ \cite{luo027001}.

\section{Dynamical spin susceptibility}

The dynamical spin susceptibility, which comes from the
particle-hole excitations, is given by
\begin{eqnarray}
 \chi ^{0}(\mathbf{q},\tau
)={\frac{1}{N}}\langle T_{\tau }S_{\mathbf{q} }^{z}(\tau
)S_{-\mathbf{q}}^{z}(0)\rangle _{0},
\end{eqnarray}
where $S_{\mathbf{q}}^{z}\equiv
{\frac{1}{2}}\sum_{\mathbf{k},\sigma }\sigma
f_{\mathbf{k+q/2},\sigma }^{\dagger }f_{\mathbf{k-q/2},\sigma }$
is the spin-density operator with $\sigma $ the spin index. Using
the transformation (\ref{unitary transformations}) and Fourier
transforming $\chi ^{0}(\mathbf{q},\tau )$ into the Matsubara
frequency space, one obtains
\begin{equation}
\chi ^{0}(\mathbf{q},i\omega_n
)=-\frac{1}{2N}{\sum_{\mathbf{k,}\mathit{ll} ^{\prime }}}^{\prime
}\nu _{\mathbf{k,}ll^{\prime }}\chi _{\mathbf{k,} ll^{\prime
}}^{0}(\mathbf{q},i\omega_n ).  \label{spin susceptibility}
\end{equation}
Here
$\nu _{\mathbf{k,}ll^{\prime }}(\mathbf{q})\equiv\left\{
1+\epsilon _{ll^{\prime }}\cos [2(\theta _{\mathbf{k}}-\theta
_{\mathbf{k+q}})]\right\}$
with $\epsilon _{ll^\prime}=1~(-1)$ for $l=l^{^{\prime }}(l\neq
l^{^{\prime }})$ and
\begin{eqnarray*}
\chi _{\mathbf{k,}ll^{\prime }}^{0}(\mathbf{q},i\omega _{n})
&=&-\frac{1}{ \beta }\sum_{i\nu
_{n}}[\mathcal{G}_{l}(\mathbf{k},i\nu _{n})\mathcal{G}
_{l^{\prime }}(\mathbf{k+q},i\nu _{n}+i\omega _{n}) \\
&&+\epsilon _{ll^{\prime }}\mathcal{F}_{l}(\mathbf{k},i\nu
_{n})\mathcal{F} _{l^{\prime }}^{\dag }(\mathbf{k+q},i\nu
_{n}+i\omega _{n})]
\end{eqnarray*}
with $\mathcal{G}_{l}$ and $\mathcal{F}_{l}$ the single-particle
normal and anomalous Green's function of band $l$. Considering the
AF vertex correction under the random-phase approximation, one
then has the renormalized spin susceptibility $\chi
(\mathbf{q},i\omega_n)=\chi ^{0}(\mathbf{q},i\omega_n)/[1+\nu
J({\bf q})\chi ^{0}(\mathbf{q},i\omega_n)]$, where $J({\bf
q})\equiv \cos (q_{x})+\cos (q_{y})$ and $\nu$ is the coupling
strength. The INS intensity, $I({\bf q},\omega)$, is proportional
to $\mathrm{Im}\chi(\mathbf{q},i\omega _{n}\rightarrow \omega
+i0^{+})$. As a matter of fact, vertex correction leads to
enhancement of the spectral intensity, but giving no qualitative
change in the lineshape.

\begin{figure}[tbp]
\vspace{-3.5cm}
\includegraphics[width=0.4\textwidth]{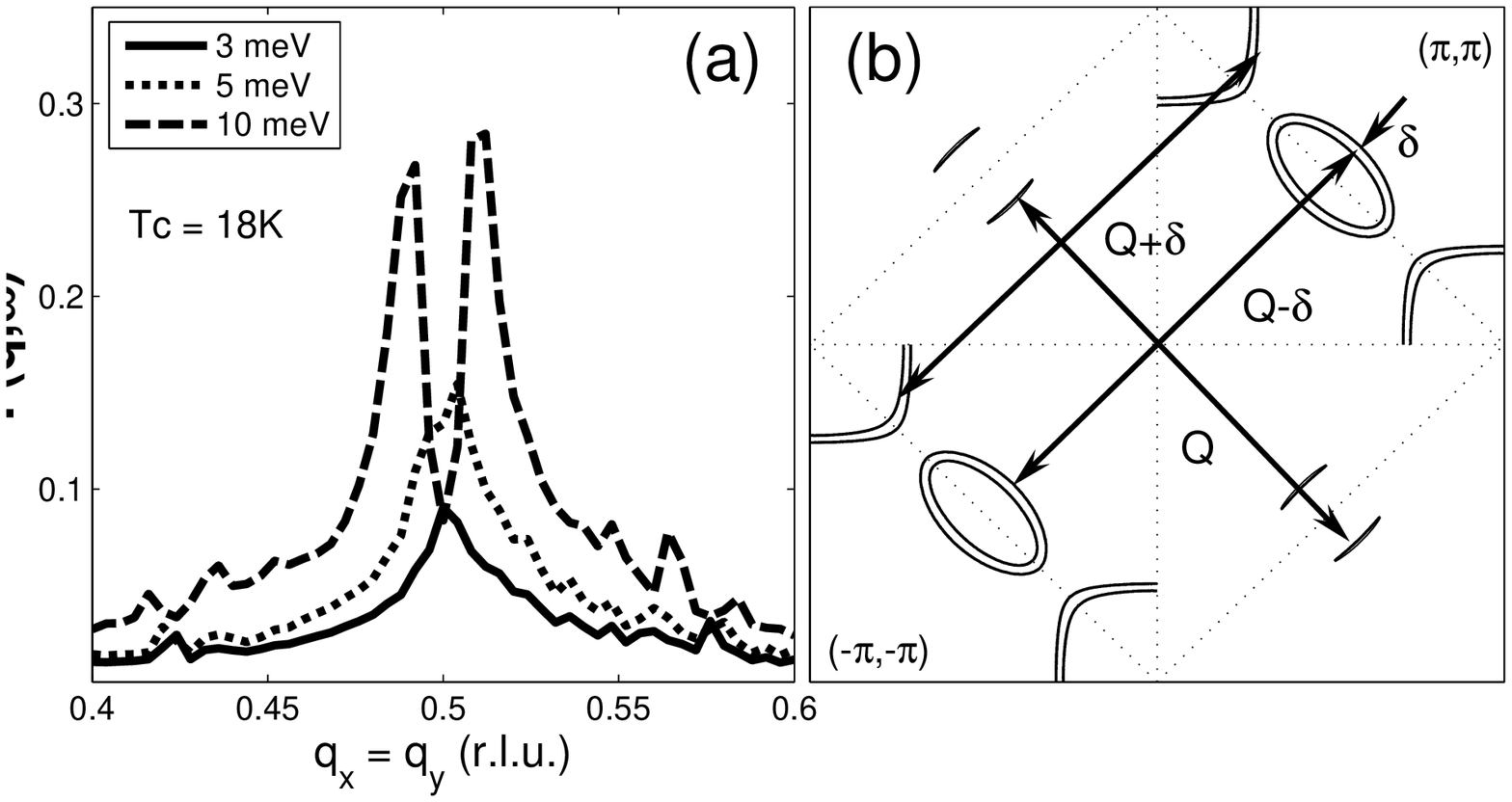}

\vspace{-3.8cm}
\includegraphics[width=0.45\textwidth]{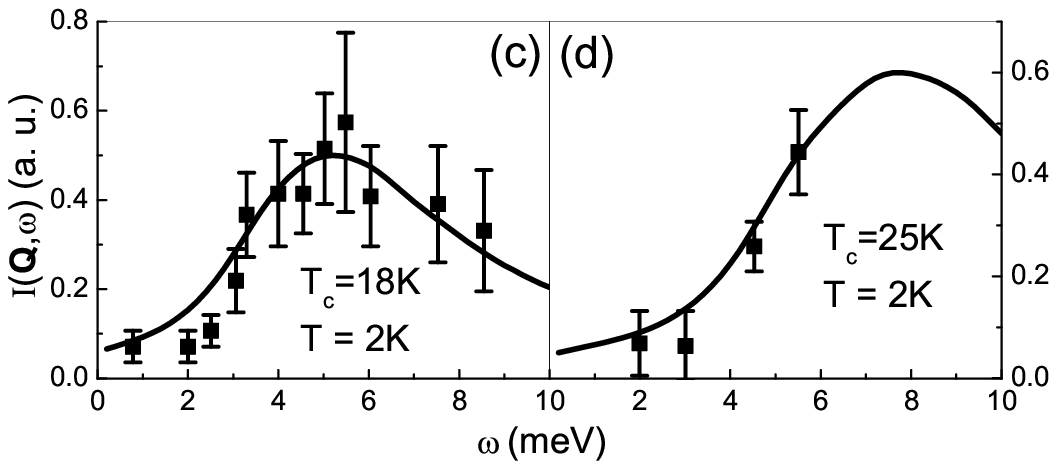}
\vspace{-0.0cm} \caption{(a) Constant $\omega$,
$\mathbf{q}$-dependent INS intensity, $I ({\bf q},
\protect\omega)$, calculated for optimally-doped ($\delta=0.15$,
$T_c=18$K) sample in the SC state ($T\rightarrow 0$). The
smearings are taken to be
$\Gamma_{\protect\alpha}=\Gamma_\beta=20$ $\mathrm{cm^{-1}}$. (b)
Energy contours of $E_{\bf k}=2.5 \mathrm{meV}$ ($5 \mathrm{meV}$)
shown in the second and fourth (first and third) quarters of the
BZ. The double-arrow lines denote the corresponding nesting
wavevectors. (c)\&(d) Comparison between theoretical calculations
(solid lines) and experimental data (dots with error bars) of
$I(\mathbf{Q}, \protect\omega)$ on SC NCCO ($\delta=0.15$). $T_c =
18$ and 25 $\mathrm{K}$ for (c) \& (d).} \label{fig2}
\end{figure}

Fig.~\ref{fig2}(a) shows the calculation of $I(\mathbf{q},\omega)
$ with different frequencies ($\omega = 3,5,10$ $\mathrm{meV}$) at
$T=0$ and optimal doping ($\delta=0.15$). The momentum is scanned
along the direction of $(0,0)\rightarrow(\protect \pi,
\protect\pi)$. For easy comparison, the coupling strength $\nu$ is
chosen to be $0.627$, same as that used in
Ref.~\cite{jianxinli2003}. The most remarkable feature is that
$I({\bf q},\omega)$ is commensurate for $\omega \leq 5$
$\mathrm{meV}$, consistent with the INS measurements of Yamada
{\em et al.} \cite{yamada137004}. In the case of higher $\omega =
10$ $\mathrm{meV}$, in contrast, $I({\bf q},\omega)$ becomes
incommensurate. The above theoretical results are in great
contrast to those obtained based on a one-band model
\cite{jianxinli2003}, where spin response is found to be
incommensurate at lower frequencies but shifted to be commensurate
at higher frequencies.

The switch from a commensurate to the incommensurate peaks upon
frequency increase can be understood in terms of the band nesting
effect. As illustrated in Fig.~\ref{fig2}(b), two sets of energy
contours: $E_{\mathbf{k}}=2.5$ $\mathrm{meV}$ and $5$
$\mathrm{meV}$ are plotted respectively in the second and fourth
and first and third quarters of the BZ. When energy is low, only
$\beta$ band opens up a contour, and the flat (nesting) portion of
the energy contours show a thin strip near the MBZ border. The
corresponding wave vector $\mathbf{q}$ (double arrow) which brings
the nesting portion into good alignment with its partner in the
other quadrant equals to $\mathbf{Q}$. Consequently, the nearly
degenerate excitations give a commensurate peak. When energy is
high, in contrast, both $\alpha$ and $\beta$ bands open up a
contour. In addition to ${\bf Q}$ nestings, the most contribution
may come from the incommensurate ${\bf Q}\pm{\bf \delta}$ nestings
[see Fig.~\ref{fig2}(b)].

The exact or near ($\delta$ is small) ${\bf Q}$ spin fluctuations
may assist the QPs to form the $d_{x^2-y^2}$-wave pairing. This is
the scenario widely believed for the hole-doped side. Based on
this pairing mechanism and taking possible ${\bf Q}$ connections
into account [in view of Fig.~2(b)], the validity of the model
pairing potential (\ref{pairV}) is justified.

One can examine the intensity at ${\bf q}={\bf Q}$ more carefully.
At low $\omega$, $I (\mathbf{Q},\omega )$ is weak, indicating that a
spin gap opens up. $I (\mathbf{Q},\omega )$ will reach its maximum
at $\omega=2\bar{\Delta}$ with a pairing-breaking gap estimated to
be $\vert \bar{\Delta}\vert ^{2}\approx\frac{1}{\phi _{\alpha }+\phi
_{\beta }}[ \int_{0}^{\phi _{\beta }}\left\vert \Delta _{\mathbf{k}
\beta }\right\vert ^{2}d\phi +\int_{\phi _{\alpha }}^{\pi
/4}\left\vert \Delta _{\mathbf{k}\alpha }\right\vert ^{2}d\phi]$.
When $\omega$ is higher, nesting portions move out of the MBZ
boundary, and consequently $I (\mathbf{Q},\omega )$ starts to
diminish. A good agreement between the theoretical calculation and
experimental $I (\mathbf{Q},\omega)$ is obtained and shown in
Fig.~\ref{fig2}(c)\&(d). In Fig.~\ref{fig2}(c) with $T_{c}=18{\rm
K}$, $\Delta _{\alpha}=22 {\mathrm{cm}^{-1}}$ and $\Delta _\beta =38
{\mathrm{cm}^{-1}}$, and $2\bar{\Delta} \simeq 5.2$ $\mathrm{meV}$
is obtained. While in Fig.~\ref{fig2}(d) with $T_{c}=25{\rm K}$,
$\Delta _{\alpha}=32 {\mathrm{cm}^{-1}}$ and $\Delta _\beta =55
{\mathrm{cm}^{-1}}$, and $2\bar{\Delta} \simeq 7.6$ $\mathrm{meV}$
is given. It is noted that the above $\Delta_\alpha$ and
$\Delta_\beta$ are taken exactly the same as those led to good fits
for the Raman scattering \cite{liu:174517}.

Recently, high-energy spin excitation of INS experiment is also
reported \cite {wilson:157001}. The energy taken in those
experiment is far above $2\Delta_l$ ($\sim$ 10 meV), beyond the
scope of the present paper. In such case, the excitation leads to
a spin-wave-like ring \cite {wilson:157001} and the effect of
pairing breaking is weak. A two-dimensional AF Heisenberg model
including nearest ($J_1$), next-nearest ($J_2$), and
next-next-nearest ($J_3$) couplings should be used to interpret
the experiments.

\section{Summary}

In summary, the gap symmetry of electron-doped cuprate
superconductors is studied based on a two-gap model. Considering a
mechanism induced by the AF spin fluctuation, a piecewise pairing
potential is proposed to account for the observed nonmonotonic
$d_{x^2-y^2}$-wave feature of the gap. Dynamical spin
susceptibility is calculated and shown to be in good agreement
with the experiment. This gives a strong support to the proposed
pairing model.

\acknowledgements

This work was supported by National Science Council of Taiwan
(Grant No. 94-2112-M-003-011) and National Natural Science
Foundation of China (Grant No. 10347149). We also acknowledge the
support from the National Center for Theoretical Sciences, Taiwan.


\end{document}